\documentclass[aps,prb,superscriptaddress,twocolumn,showpacs]{revtex4-2}
\usepackage{amsmath}
\usepackage{amssymb}
\usepackage[pdftex]{graphicx,color}
\usepackage{bm}
\usepackage{cases}
\usepackage{multirow}
\usepackage{braket}
\usepackage{mathtools}
\usepackage{comment}
\usepackage[pdftex]{hyperref}
\usepackage[newcommands]{ragged2e}

\def\stacksymbols #1#2#3#4{\def\theguybelow{#2}
    \def\verticalposition{\lower#3pt}
    \def\spacingwithinsymbol{\baselineskip0pt\lineskip#4pt}
    \mathrel{\mathpalette\intermediary#1}}
\def\intermediary#1#2{\verticalposition\vbox{\spacingwithinsymbol
      \everycr={}\tabskip0pt
      \halign{$\mathsurround0pt#1\hfil##\hfil$\crcr#2\crcr
               \theguybelow\crcr}}}

\begin{document}
\title{Bulk-Edge Correspondence for Point-Gap Topological Phases in Junction Systems}

\author{Geonhwi Hwang}
\affiliation{Department of Applied Physics, Hokkaido University, Sapporo 060-8628, Japan}
\author{Hideaki Obuse}
\affiliation{Department of Applied Physics, Hokkaido University, Sapporo 060-8628, Japan}
\affiliation{Institute of Industrial Science, The University of Tokyo, 5-1-5 Kashiwanoha, Kashiwa, Chiba 277-8574, Japan}

\begin{abstract}
The bulk-edge correspondence is one of the most important ingredients in the theory of topological phases of matter.
While the bulk-edge correspondence is applicable for Hermitian junction systems where two subsystems with independent topological invariants are connected to each other, it has not been discussed for junction systems with non-Hermitian point-gap topological phases.
In this Letter, based on analytical results obtained by the extension of non-Bloch band theory to junction systems, we establish the bulk-edge correspondence for point-gap topological phases in junction systems.
We also confirm that almost all the eigenstates are localized near the interface which are called the ``{\it non-Hermitian proximity effects}''. One of the unique properties is that the localization length becomes the same for both subsystems nevertheless those model-parameters are different.
\end{abstract}

\maketitle
{\it Introduction}. Non-Hermitian systems have recently received a lot of attention since they possess novel physical phenomena and richer topological properties than Hermitian systems\cite{Hatano1996,Hatano1997,Hatano1998,Bender1998,Bender2002,Rudner2009,Esaki2011,Hu2011,Lee2014,Lee2014_2,Lee2016,Mochizuki2016,Gonzalez2017,Leykam2017,Xiao2017,Kawabata2018,Wang2018,Nakagawa2018,Yao2018,Yao2018_2,Gong2018,Lieu2018,Zyuzin2018,Yoshida2018,Chen2018,Shen2018,Molina2018,Takata2018,Zhou2018,Zhou2018_2,Kawabata2019,Yokomizo2019,Imura2019,Liu2019,Yamamoto2019,Xiao2019,Li2019,Ezawa2019,Okugawa2019,Budich2019,Yang2019,Yoshida2019,Wu2019,Ashida2020,Slager2020,Okuma2020,Zhang2020,Weidemann2020,Xiao2020,Bergholtz2021,Hatano2021,Longhi2021,Longhi2022,Kawasaki2022,Okuma2023,Schindler2023}.  In particular, non-Hermitian physics describes not only open quantum systems but also dissipative classical systems on equal footing, due to mathematical similarity of the fundamental equations of motion in both 
 systems\cite{Makris2008,Klaiman2008,Guo2009,Ruter2010,Feng2013,Zeuner2015,Malzard2015,Rosenthal2018,Parto2018,Malzard2018}.

A topological phase refers to a state of matter with a non-trivial topological invariant for energy-gapped states. There are two types of energy gaps defined in the non-Hermitian system, namely point-gaps and line-gaps\cite{Gong2018,Kawabata2019}. Hereafter, we focus on topological phases originating with the point-gap (point-gap topological phases, in short) which is unique to the non-Hermitian systems\cite{Yao2018,Yokomizo2019}.

The bulk-edge correspondence (BEC) for the point-gap topological phases has been studied\cite{Slager2020,Okuma2020,Zhang2020} and the following statements are confirmed for systems without any symmetry in one dimension systems: A spectrum for a system with periodic boundary conditions (PBC) forms closed curve(s) winding a point on the complex plane, giving a topological invariant which is called a winding number for that point. Then, the spectrum for the corresponding system with semi-infinite boundary conditions (SIBC) is equal to the the spectrum for the system with PBC (PBC spectrum, in short) together with the area, which is the set of points for which the winding number is non-trivial. The spectrum for the corresponding system with open boundary conditions (OBC) forms non-closed curve(s) and appears on the SIBC spectrum. Further, it has been revealed that the point-gap topological phases give rise to the skin effect which makes all eigenstates localized near the open boundaries.

As mentioned above, the BEC for the point-gap topological phases has only been discussed with PBC, SIBC and OBC so far. 
While interface states appearing at the region where two subsystems with different point-gap topological phases are connected have been studied\cite{Weidemann2020,Xiao2020,Longhi2021,Longhi2022}, the BEC for the junction geometry has not been clarified, nevertheless the BEC has been established even for junction systems in the Hermitian system.

In this Letter, we extend the concept for the BEC for the point-gap topological phases in non-Hermitian systems to junction systems. To this end, we consider a one-dimensional junction system with PBC where two ends of a subsystem are connected to those of the other subsystem so that the whole system forms a ring geometry. Here, each subsystem has asymmetric hopping terms and its own point-gap topological phase. We confirm that the spectrum for the junction system with PBC appears where the winding number for each subsystem is different. We further study the eigenstates in the junction systems and find that the almost all the eigenstates are localized near the interface, which are called the ``non-Hermitian proximity effects''. This establishes the BEC for the point-gap topological phases in junction systems with PBC. We also discuss the junction system with OBC where one end of a subsystem is connected to that of the other subsystem so that open boundaries exist at both ends of the wh
 ole system. We confirm that the spectrum for the junction system with OBC appears where the winding number for the corresponding junction system with PBC is non-trivial, revealing that the existing BEC for the point-gap topological phases\cite{Gong2018,Okuma2020,Zhang2020} can be applied to the junction systems with OBC as well.
%

{\it Model}. We start with a one-dimensional tight-binding model with asymmetric hopping terms, so-called Hatano-Nelson model\cite{Hatano1996,Hatano1997,Hatano1998} whose Hamiltonian is 
%
\begin{equation}
\label{21}
H=\sum_{n} \left( t^{+} c_{n+1}^{\dagger} c_n +t^{-} c_n^{\dagger} c_{n+1} +  \epsilon c_n^{\dagger} c_n  \right) ,
\end{equation}
\noindent where $t^{\pm} \coloneqq te^{\pm \gamma}$ for $t, \gamma \in \mathbb{R} $ and $\epsilon \in \mathbb{R}$ which corresponds to the on-site potential.
By applying the Fourier transform to Eq.\,(\ref{21}), the PBC spectrum $\sigma_{\rm PBC}$ for this Hamiltonian is given by
%
\begin{equation}
\label{22}
\sigma_{\rm PBC} = \{ 2t\cos \left(k-i\gamma\right) +\epsilon \, | \, k\in \left[0 , 2\pi\right) \} .
\end{equation}
Then, all the eigenenergies lie on an ellipse centered at $\epsilon$ on the complex plane, and the point-gap is open for all points surrounded by the ellipse. 
The topological invariant for the point-gap topological phases at a point $E_p$, can be defined as winding number $w \left(E_p\right)$ as follows\cite{Gong2018,Kawabata2019}:
%
\begin{align}
\label{23}
w \left(E_p\right) &\coloneqq  \frac{1}{2\pi i} \int_0^{2\pi} dk \partial_k\ln {\rm det} \left( H(k) -E_p \right) ,
\end{align}
where $H(k)$ is the momentum representation of Eq.\,(\ref{21}). Eq.\,(\ref{23}) means how many times the spectrum $\sigma_{\rm PBC}$ winds around the reference point $E_p$ on the complex plane.
The BEC for the OBC spectrum $\sigma_{\rm OBC}$, as mentioned in introduction, can be written down as follows:
%
\begin{equation}
\label{24}
\sigma_{\rm OBC} \subset \{ E_p \vert w (E_p) \neq 0 \lor E_p \in \sigma_{\rm PBC}  \} .
\end{equation}

To extend the above mentioned BEC to junction systems with PBC, we consider a ring geometry where two subsystems with asymmetric hoppings, subsystem ${\rm I}$ and ${\rm II}$, have independent parameters. First, we define the Hamiltonian for the whole system as follows:
%
\begin{align}
\label{25}
H=H_1 (1,N_1 ) + H_2 (N_1 +1 , N_2-1)+H_{\rm BC}.
\end{align}
Here, $H_{i=1,2} (N ,M)$ which corresponds to the subsystem ${\rm I}$ and ${\rm II}$ is given by
%
\begin{align}
\label{26}
H_i (N ,M)=\sum_{n=N }^{N + M-1} \left( t_i^{+} c_{n+1}^{\dagger} c_n + t_i^{-} c_n^{\dagger} c_{n+1} + \epsilon_i c_n^{\dagger} c_n  \right) ,
\end{align}
in real space, where $t_i^{\pm} \coloneqq t_i e^{\pm \gamma_i}$, $t_i , \gamma_i , \epsilon_i \in \mathbb{R} $.
$H_{\rm BC}$ determines PBC or OBC.
For the sake of simplicity, hereafter, we assume $t_i > 0$ and $\gamma_1 +\gamma_2>0$. 

To discuss the BEC for the junction system with PBC, we decouple the two subsystems, $H_1$ and $H_2$, and impose PBC on each subsystem.
Then, we obtain the PBC spectrum for each subsystem, $\sigma_{\rm PBC}^{(i=1,2)}$, from Eq.\,(\ref{22}). Applying Eq.\ (\ref{23}), the winding number for each subsystem, $w_{i=1,2}$, is given by the sign of $\gamma_i$ for the reference point $E_P$ located inside $\sigma_{\rm PBC}^{(i)}$. We also introduce the following winding number, $W_i \coloneqq w_i(\epsilon_i)={\rm sgn} \left(\gamma_i\right)$ where $\epsilon_i$ means the center of $\sigma_{\rm PBC}^{(i)}$, which we use hereafter.
%
%

{\it Junction systems with PBC.} Here, we analytically solve the eigenvalue $E(\in\mathbb{C})$ and the (right) eigenvectors $\ket{\psi}$ of the the Schr\"odinger equation
%
\begin{equation}
\label{31}
H \ket{\psi} = E \ket{\psi} ,\indent \ket{\psi} = \left(\psi_1 , \cdots , \psi_{N_1+N_2}\right)^T ,
\end{equation}
for the Hamiltonian of the junction system with PBC in Eq.\,({\ref{25}) with $H_\text{BC}=t_2^+ c_{1}^\dagger c_{N_1+N_2} + t_2^-c_{N_1+N_2}^\dagger c_{1}+\epsilon_2c_{N_1+N_2}^\dagger c_{N_1+N_2}$. Our derivation is based on the extension of the non-Bloch band theory\cite{Yao2018,Yokomizo2019} to junction systems (see the Supplemental Material for the details of the derivation\cite{supplement}).
We obtain two recurrence relations for the bulk region as 
%
\begin{align}
\label{32}
E_1\psi_n &= t_1^{+} \psi_{n-1} +  t_1^{-} \psi_{n+1}  \quad \left(n\in [2,  N_1-1] \right) ,\\
\label{33}
E_2\psi_n &= t_2^{+} \psi_{n-1} +  t_2^{-} \psi_{n+1} \quad \left(n\in [N_1 +2 , N_1+N_2-1] \right) ,
\end{align}
where $E_1 \coloneqq E- \epsilon_1 $ and $E_2 \coloneqq E- \epsilon_2 $. For $n=1, N_1, N_1+1,$ and $N_1+N_2$, we obtain four boundary conditions:
%
\begin{align}
\label{b1}
&E_1 \psi_1 = t_2^{+} \psi_{N_1 + N_2 } +  t_1^{-} \psi_{2} ,\\
\label{b2}
&E_1 \psi_{N_1} = t_1^{+} \psi_{N_1 -1} +  t_1^{-} \psi_{N_1 +1} ,\\
\label{b3}
&E_2 \psi_{N_1 +1} = t_1^{+} \psi_{N_1 } +  t_2^{-} \psi_{N_1 +2} ,\\
\label{b4}
&E_2 \psi_{N_1 + N_2 } = t_2^{+} \psi_{N_1 -1 } +  t_2^{-} \psi_{1} .
\end{align}
Here, without loss of generality, we represent
%
\begin{equation}
\label{34}
E_1 = t_1 \left( x_1 + x_1^{-1} \right), \quad E_2 = t_2 \left( x_2 + x_2^{-1} \right), 
\end{equation}
with $x_1 , x_2 \in \mathbb{C}$ whose absolute values belong to $(0,1]$.
Then, the general solution is given by
%
\begin{equation}
\label{35}
\psi_n = 
\left\{ \begin{array}{ll}
\phi_1 \left(e^{\gamma_1}x_1 \right)^n + \phi_2 \left(e^{\gamma_1}/x_1 \right)^n  & \left( n \in [1, N_1] \right) ,
\\
\phi_3 \left(e^{\gamma_2}x_2 \right)^{n-N_1} + \phi_4 \left(e^{\gamma_2}/x_2 \right)^{n-N_1} & \left( n\in [N_1 +1,N_1 +N_2]  \right) ,
\end{array} \right. 
\end{equation}
where $\phi_{i=1,2,3,4}$ are constants.
By substituting Eq.\,(\ref{35}) into Eqs.\,(\ref{b1})-(\ref{b4}) and examining non-trivial $\phi_{i}$, we can determine the values of $x_1$ and $x_2$.
Similar to the non-Bloch band theory, there should be $N_1+N_2$ pairs $(x_1^{(m)},x_2^{(m)})$ for $m=1,\cdots ,N_1+N_2$ so that the spectrum becomes continuous when $N_1+N_2\rightarrow \infty$.
We can express the $m$th eigenenergy $E^{(m)}$ and the corresponding eigenfunction $\psi_n^{(m)}$ exactly by substituting $x_1^{(m)}$ (or $x_2^{(m)}$) into Eqs.\,(\ref{34}) and (\ref{35}), respectively.
With the eigenenergy $E^{(m)}$(see Sec.\ II in Ref.\,\cite{supplement} for details), we establish the BEC for the point-gap topological phases in junction systems with PBC as follows:
%
\begin{align}
\label{38}
\sigma_{\rm PBC}^{\rm junc} \subset \{ E_p \vert \Delta w (E_p) \neq 0 \lor E_p \in (\sigma_{\rm PBC}^{(1)} \cup \sigma_{\rm PBC}^{(2)}) \} ,\\
\label{38_2}
\Delta w (E_p) \coloneqq |w_1(E_p)-w_2(E_p)|. \hspace{31pt} 
\end{align}
In addition, the eigenfunction can be approximated as
%
\begin{equation}
\label{36}
\left| \psi_n \right| \propto 
\left\{ \begin{array}{ll}
e^{\kappa_1 n} & \quad (n\in[1,N_1]),
\\
e^{\kappa_2 n} & \quad (n \in [N_1+1,N_1+N_1]),
\end{array} \right.
\end{equation}
where
%
%
\begin{align}
\label{37}
\kappa_i = 
\left\{ \begin{array}{ll}
\gamma_i - l_i & \quad (l_i  \leq \gamma_1+\gamma_2),
\\
\gamma_i - {\rm sgn} \left(\gamma_{j\neq i}\right) l_i & \quad (l_i >\gamma_1+\gamma_2) ,
\end{array} \right.
\\
\label{37_2}
l_i \coloneqq -\log \vert x_i \vert \quad (i=1,2) . \hspace{29.5pt}
\end{align}
From Eqs.\,(\ref{36}) and (\ref{37}), we find that almost all the eigenstates are localized in junction system with PBC.
For the eigenenergies corresponding to the delocalized eigenstates, $\gamma_1$ and $\gamma_2$ must be positive so that $l_i=\gamma_i$ for $i=1,2$.
These eigenenergies appear at the intersections of the PBC spectra for each subsystem.
These are the main results of this Letter.

The statements above can be regarded as a natural extension of the BEC for juction systems in Hermitian systems.
Below, we will examine several cases to validate our results and discuss the BEC for the point-gap topological phase in junction systems with PBC.
%
%
%
%

{\it Case I}: $W_1=W_2$. 
\begin{figure}
  \centering
    \includegraphics[width=0.49\linewidth]{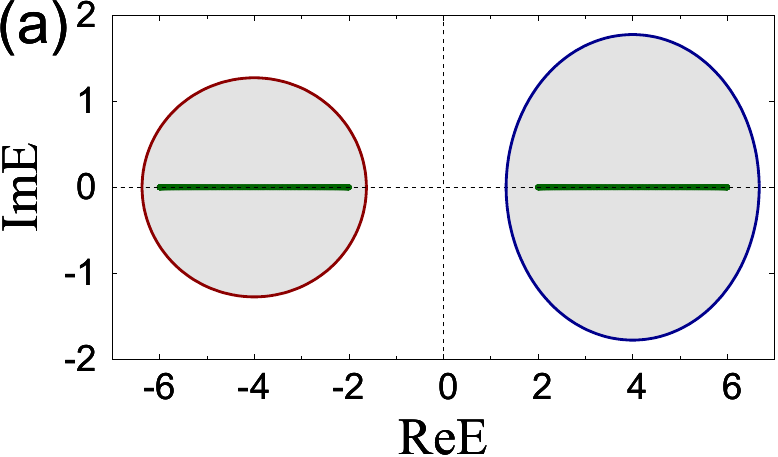}
    \includegraphics[width=0.49\linewidth]{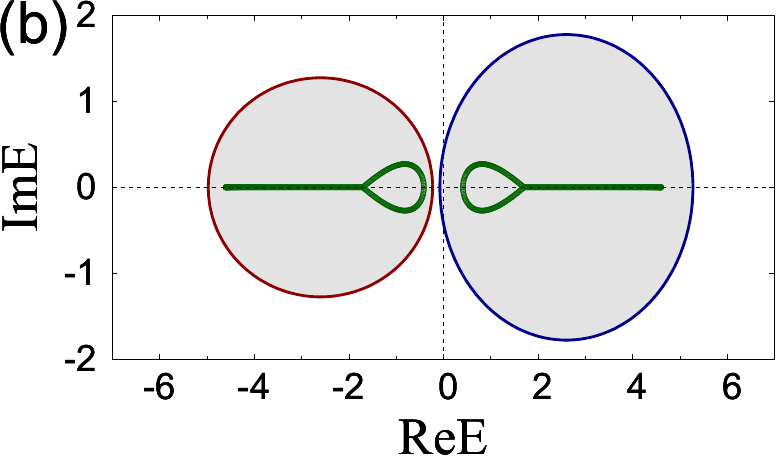}
    \includegraphics[width=0.49\linewidth]{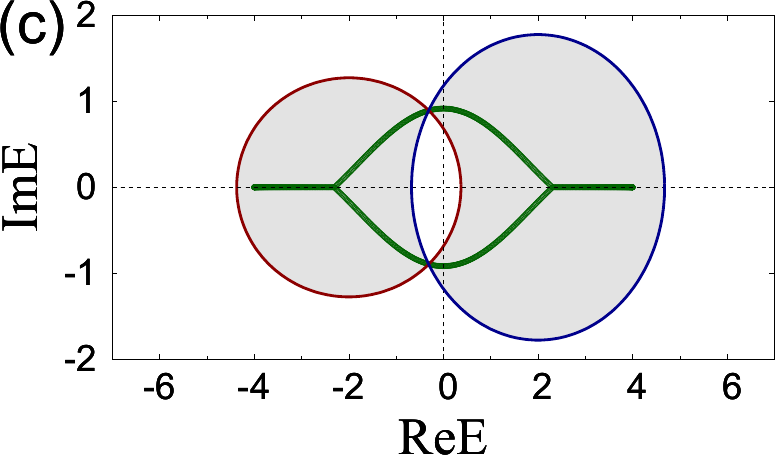}
    \includegraphics[width=0.49\linewidth]{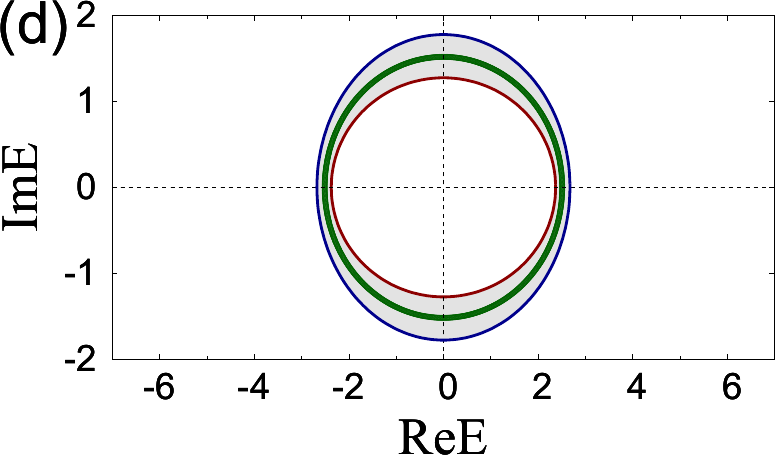}
\caption{PBC junction spectra (green dots) for different on-site potentials.
(a)$\epsilon_2 =4$, (b)$\epsilon_2 =2.6$, (c)$\epsilon_2 =2$ and (d)$\epsilon_2 =0$ with the condition $\epsilon_1=-\epsilon_2$.
The other parameters are set as $t_1=t_2=1$, $\gamma_1=0.6$, $\gamma_2=0.8$ and $N_1=N_2=500$.
PBC spectra for subsystems {\rm I} and {\rm II}, $\sigma_{\rm PBC}^{(1)}$ and $\sigma_{\rm PBC}^{(2)}$, are shown by the red and blue ellipses, respectively.
The gray regions mean $\Delta w(E_p) \neq 0$.}
\label{f1}
\end{figure}
Figure\,\ref{f1} shows the spectra for the junction systems with PBC (PBC junction spectra, in short) on the complex plane for different values of on-site potentials.
Since $\gamma_1$ and $\gamma_2$ are set to be positive, $W_1=W_2=1$.
As $\epsilon_2$ approaches infinity, each subsystem becomes isolated from each other in the energy space, expected to be independent Hatano-Nelson model, Eq.\,(\ref{21}), with OBC. Then, the PBC junction spectrum forms two energy bands on the real axis, which are the same with two spectra of the Hatano-Nelson model with OBC with $\epsilon_1$ and $\epsilon_2$.
The numerical result in Fig.\,\ref{f1}(a) agrees with this expectation even at $|\epsilon_1-\epsilon_2|=8$.

We check how the PBC junction spectrum behaves as we decrease the difference in on-site potential $|\epsilon_1 -\epsilon_2|$. As shown in Fig.\,\ref{f1}(b), when the PBC spectra for the subsystems {\rm I} and {\rm II} get closer but are not overlapped each other, we see that the PBC junction spectrum forms a loop at the edge of each band
(note that this does not imply the inconsistency of the BEC for the point-gap topological phases\cite{Okuma2020} since the system is subject to PBC.).
When $\sigma_{\rm PBC}^{(1)}$ and $\sigma_{\rm PBC}^{(2)}$ begin to intersect, the two loops merge into a single loop passing through the crossing points [Fig.\,\ref{f1}(c)]. 
When the on-site potentials become equal to each other, the PBC junction spectrum becomes an ellipse, similar to the PBC spectrum in Eq.\,(\ref{22}), as shown in Fig.\,\ref{f1}(d).
For all cases, we observe that the PBC junction spectrum $\sigma_{\rm PBC}^{\rm junc}$ appears in the region where $\Delta w \neq 0$, satisfying Eq.\,(\ref{38}).
\begin{figure}
\centering
\begin{minipage}{0.36\linewidth}
       \includegraphics[width=\linewidth]{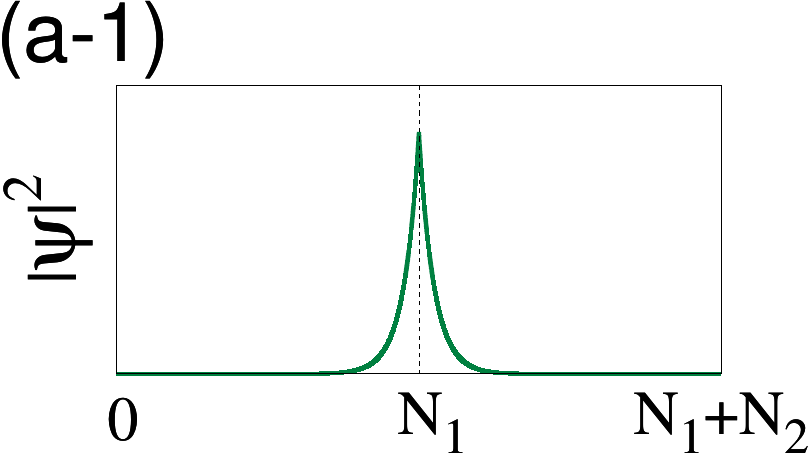}
       \includegraphics[width=\linewidth]{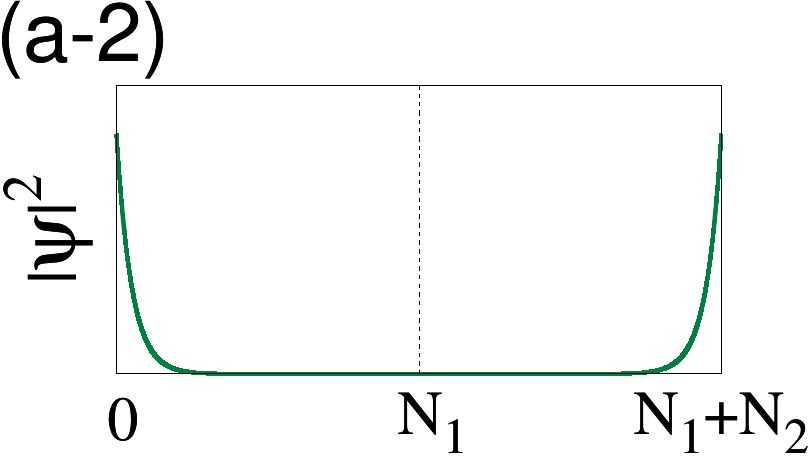}
\end{minipage}
\begin{minipage}{0.62\linewidth}
\vspace*{\fill} %
\centering
\includegraphics[width=\linewidth]{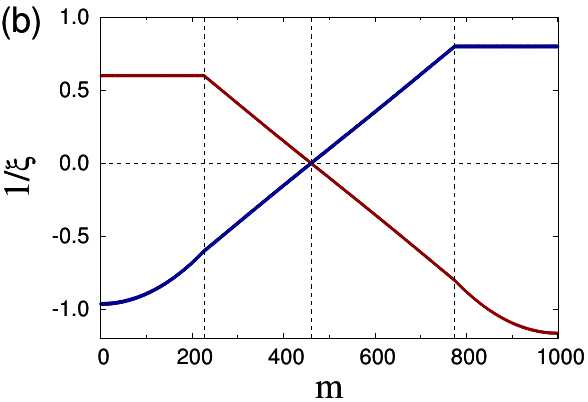}
\end{minipage}
\caption{(a): Schematic plots of eigenstates $|\psi_n |^2$ corresponding to complex eigenenergies appearing inside $\sigma_{\rm PBC}^{(1)}$ [(a-1)] and $\sigma_{\rm PBC}^{(2)}$ [(a-2)] in Fig.\,\ref{f1}(c).
The eigenstates are localized near the boundary of subsystems.
Note that PBC are imposed on both edges.
(b): Inverse of the localization lengths $\xi_{i=1}$ (red dots) and $\xi_{i=2}$ (blue dots) in the subsystem {\rm I} and {\rm II}, respectively, in the junction system of Fig.\,\ref{f1}(c).
$m$ is numbered in ascending order of the real part of the eigenenergy values.}
\label{f2}
\end{figure}

Next, we consider the probability distribution function (PDF) of an eigenstate $|\psi_n|^2$ in junction systems with PBC.
We show two PDFs corresponding to the eigenenergies in Fig.\,\ref{f1}(c) as typical examples.
We see that both PDFs are localized near the boundaries of the subsystems.
We remark that the eigenstate in Fig.\,\ref{f2}(a-1) [Fig.\,\ref{f2}(a-2)] whose eigenenergy is inside the $\sigma_{\rm PBC}^{(1)}$ ($\sigma_{\rm PBC}^{(2)}$) is localized only near $n=N_1$ ($n=1$).
This can be explained as follows.
Since the subsystem ${\rm I}$ dominates the eigenstate in Fig.\,\ref{f2}(a-1), the PDF shows the peak near the right edge of the subsystem ${\rm I}$ as the skin effect due to $\gamma_1 >0$.
Meanwhile, for the subsystem ${\rm II}$, the PDF localized near the left edge of the subsystem ${\rm II}$ as the proximity effects of the peak in the subsystem ${\rm I}$.
We shall henceforth call this the {\it non-Hermitian proximity effect}.
The result in Fig.\,\ref{f2}(a-2) can be explained in the same way as above.

To study the localization properties further, we calculate the localization length for each subsystem, $\xi_{i=1,2}$, defined as $\vert \psi_n \vert \propto e^{n/\xi_i}$ for $n$ in each subsystem, by numerical fittings.
Note that $\xi_i$ can take a negative value representing the exponential decay with increasing $n$.
Fig.\,\ref{f2}(b) shows $1/\xi_i$ of all the eigenstates in Fig.\,\ref{f1}(c).

First, we consider the eigenenergies on the real axis.
We see that $\xi_1=1/\gamma_1$ for all negative eigenenergies on the real axis, which is consistent with the localization length of an isolated subsystem {\rm I} with OBC, while the other localization length, $\xi_2$, increases gradually as the corresponding eigenenergy decreases ($m \leq 227$).
According to our analytic calculations, the localization length $\xi_2$ whose eigenenergy is negative infinity as $\epsilon_1=-\epsilon_2 \rightarrow \infty$, eventually converges to zero, which is reasonable by considering the physical meaning.
The above analysis can also be applied to all positive eigenenergies on the real axis ($m \geq 774$).

Next, we shift our focus to the complex eigenenergies ($227<m<774$).
In this region, $\xi_1=-\xi_2$ is expected by our analytic results \cite{supplement} and confirmed numerically in Fig.\,\ref{f2}(b).
Remarkably, the absolute values of the localization lengths are exactly the same, nevertheless the values of the parameters $\gamma_{i}$ and $\epsilon_{i}$ are different for each subsystem. 
According to our analytic calculation (Sec. II A in \cite{supplement}), $\vert \xi_1\vert=\vert \xi_2\vert$ remains satisfied even when $t_{1} \ne t_2$.
This is one of the unique properties of non-Hermitian proximity effects.
Further, we concentrate on the eigenenrgies at the intersections of $\sigma_{\rm PBC}^{(1)}$ and $\sigma_{\rm PBC}^{(2)}$ in Fig.\,\ref{f1}(c), where the winding number cannot be defined.
The localization lengths of the corresponding eigenenergies are shown at $m=460,461$ in Fig.\,\ref{f2}(b).
Since $1/\xi_{i=1,2}=0$, we find the eigenstates for these two eigenenergies delocalize.
This result also agrees with our analytic calculations.

Our investigation confirms that the results for additional cases in Fig.\,\ref{f1} are consistent with our analytic results.
We consider that the BEC we established can generally be applied to the point-gap topological phases in junction systems.
%
%

{\it Case II}: $W_1\neq W_2$. 
\begin{figure}
  \centering
\includegraphics[width=0.49\linewidth]{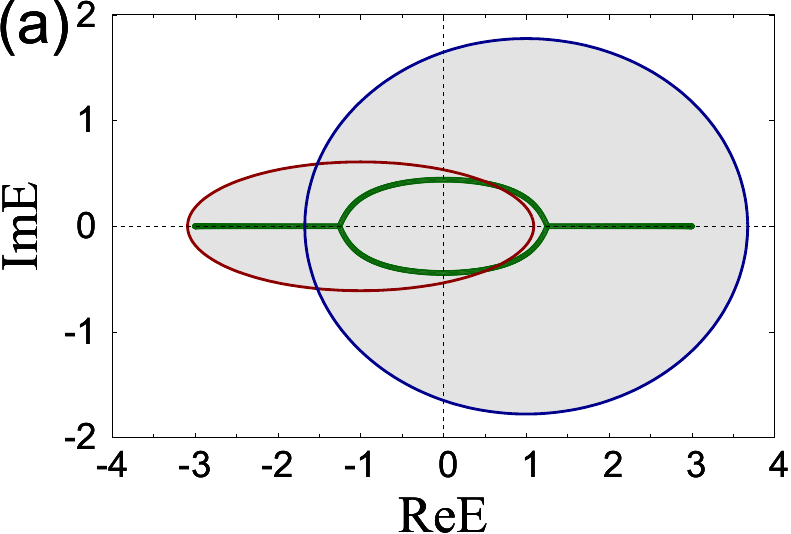}
\includegraphics[width=0.49\linewidth]{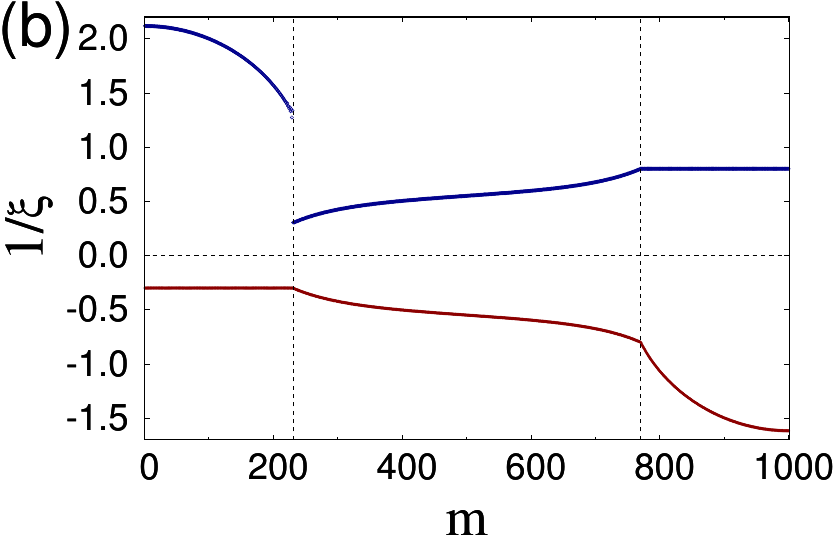}
  \caption{(a): PBC junction spectrum (green dots).
All the parameters are the same with Fig,\,\ref{f1}(c), but only the value of $\gamma_1$ is set as $-0.3$.
PBC spectra for subsystems {\rm I} and {\rm II}, $\sigma_{\rm PBC}^{(1)}$ and $\sigma_{\rm PBC}^{(2)}$, are shown by the red and blue ellipses, respectively.
The gray regions mean $\Delta w \neq 0$.
(b): Inverse of the localization lengths $\xi_{i=1}$ (red dots) and $\xi_{i=2}$ (blue dots) in the subsystem {\rm I} and {\rm II}, respectively, in the junction system of Fig.\,\ref{f3}(a).
$m$ is numbered in ascending order of the real part of the eigenenergy values.}
  \label{f3}
\end{figure}
Figure\,\ref{f3}(a) shows the PBC junction spectrum where $W_1=-1$ and $W_2=1$. 
In this case, we see that the spectrum appears even in the internal area shared by $\sigma_{\rm PBC}^{(1)}$ and $\sigma_{\rm PBC}^{(2)}$ while it does not where $W_1=W_2$ [Fig.\,\ref{f1}(c) and (d)].
But this is not a violation of BEC for junction systems with PBC since $\Delta w =2$ for that area.
Therefore, Eq.\,(\ref{38}) is also confirmed here without exception.
Also, the eigenstates, as can be seen in Fig.\,\ref{f3}(b), are localized near the boundaries of the subsystems, exhibiting the non-Hermitian proximity effects.
The localization lengths in Fig.\,\ref{f3}(b) show the same behaviors as the previous case [Fig.\,\ref{f2}(b)].
%
%
%
%
%
%

{\it Junction systems with OBC}. To clarify of the BEC for the point-gap topological phases in junction systems, hereafter, we consider the junction system with OBC in Eq.\,(\ref{25}) with $H_\text{BC}=\epsilon_2 c_{N_1+N_2}^\dagger c_{N_1+N_2}$.
Since OBC is implemented by removing hopping terms between two neighboring sites,
there are $N_1+N_2$ cases for implementing the removal of hopping terms in the model. 
While, in principle, our analytical method can be applied to the other cases, we focus on the above $H_\text{BC}$ for the present analytical calculation.

The eigenfunction can be approximated as (see Ref.\,\cite{supplement})
%
%
\begin{equation}
\label{20}
\left| \psi_n \right| \propto 
\left\{ \begin{array}{ll}
\exp[(\gamma_1 + l_1 ) n] &\quad (n \in [1,N_1]),
\\
\exp[(\gamma_2 - l_2 ) n] &\quad (n \in [N_1+1,N_1+N_2]).
\end{array} \right. 
\end{equation}
We find that there is no solution where $l_1, l_2 >0$.
This means that the spectrum for the junction system with OBC must appear on the real axis, particularly on the OBC spectra for each subsystem.
\begin{figure}
  \centering
\includegraphics[width=0.48\linewidth]{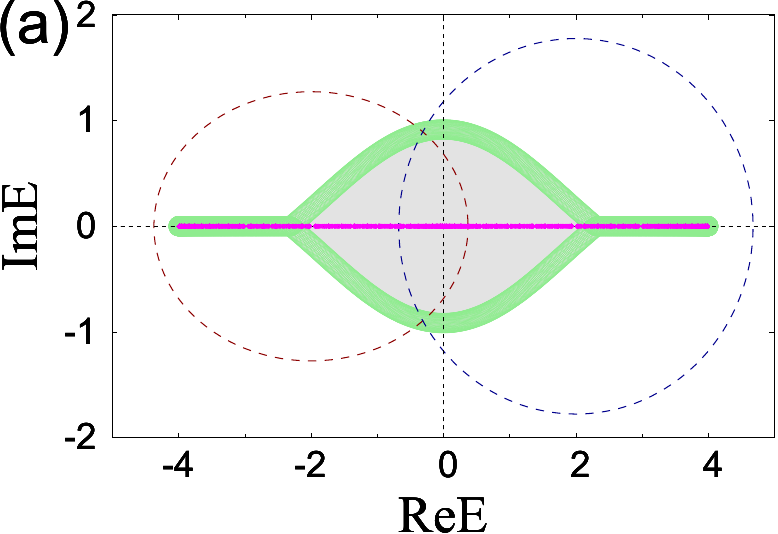}
\includegraphics[width=0.50\linewidth]{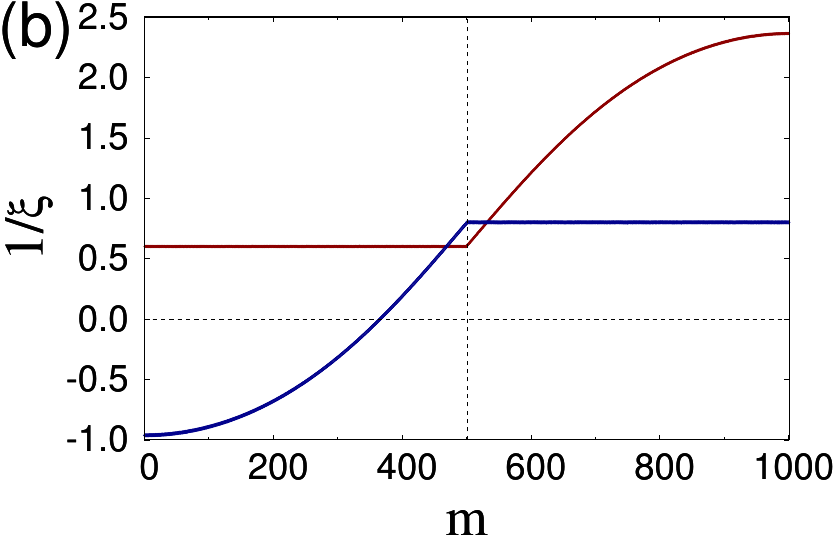}
  \caption{(a): Spectra (magenta dots) of the junction system with all possible OBC ($N_1+N_2$ cases).
All the parameters are the same with Fig,\,\ref{f1}(c) except for the system size ($N_1=N_2=30$).
PBC spectra for subsystems {\rm I} and {\rm II}, $\sigma_{\rm PBC}^{(1)}$ and $\sigma_{\rm PBC}^{(2)}$, are shown by the red and blue ellipses, respectively.
The PBC junction spectrum is shown by the green dots and the gray region means $w^{\rm junc} \neq 0$.
(b): Inverse of the localization lengths $\xi_{i=1}$ (red dots) and $\xi_{i=2}$ (blue dots) in the subsystem {\rm I} and {\rm II}, respectively, in the junction system with OBC.
All the parameters are the same with Fig.\,\ref{f1}(c). $m$ is numbered in ascending order of the real part of the eigenenergy values.}
  \label{f4}
\end{figure}

Figure\,\ref{f4}(a) shows the spectra for the junction system with all possible OBC calculated numerically.
As discussed above, we numerically confirm that all the spectra appear on the real axis.
Further, regardless of the removal position, it can be numerically confirmed and analytically explained that all spectra appear on and inside $\sigma_{\rm PBC}^{\rm junc}$ without exception.
Actually, the winding number of the junction system $w^{\rm junc}(E_p )$ which is not well defined by Eq.\ (\ref{23}) is estimated to be unity since a Hamiltonian connected by the continuous deformation of $t_1=t_2$ and $\gamma_1=\gamma_2>0$ without closing the point gap gives $w(E_p) = 1$ from Eq.\ (\ref{23}), where $E_p$ locates in the point gap.
Thus, we confirm that the spectrum for the junction system with OBC (OBC junction spectrum, in short) appears on and inside the spectrum for the corresponding junction system with PBC regardless of the removal position for hopping terms.
This result is consistent with the BEC for the point-gap topological phases\cite{Okuma2020}.

Figure\,\ref{f4}(b) shows the localization properties of the junction system with OBC which is implemented by removing the hopping terms between $n=1$ and $n=N_1+N_2$.
We see that $\xi_1=1/\gamma_1$ ($\xi_2=1/\gamma_2$) for all the eigenenergies on the OBC spectrum for the subsystem ${\rm I}$ (${\rm II}$), which is also consistent with localization length of an isolated subsystem ${\rm I}$ (${\rm II}$).
In contrast to the previous arguments on Figs.\,\ref{f2}(b) and \ref{f3}(b), we see that both $\xi_1$ and $\xi_2$ are positive for $m \geq 367$, meaning the correspponding PDFs show the peak only at the right edge of the whole system, not exhibiting the non-Hermitian proximity effects.
This result is reasonable because of the absence of the hopping between $n=1$ and $N_1+N_2$ and $\gamma_1,\gamma_2>0$.
For the eigenenergies whose eigenstates are dominated by the subsystem ${\rm I}$ ($m\leq 366$), however, we find the non-Hermitian proximity effects even in the junction system with OBC.
%
%

{\it Conclusion}. In this Letter, we have established the BEC for the point-gap topological phases in junction systems.
To summarize, for the point-gap topological phases in junction systems with PBC, the PBC junction spectra do not appear where the winding number for each subsystem is equal.
Further, almost all the eigenstates are localized near the interface and exhibit the non-Hermitian proximity effects.
We also revealed that the OBC junction spectrum appears on and inside the corresponding PBC junction spectrum.
Thereby, the BEC for the point-gap topological phases \cite{Gong2018,Okuma2020,Zhang2020} can be applied to the junction systems with OBC as well.

Since the BEC for junction systems in Hermitian systems requiring that the number of edge states is given by the difference in topological number for each subsystem is generally valid, the BEC for non-Hermitian junction systems we established, Eqs. (\ref{38}) and (\ref{38_2}), is regarded as a natural extension of the BEC for Hermitian junction systems.
Therefore, while our conclusion is derived from a specific lattice model, we consider that the present statements can be applied to more general junction systems with point-gap topological phases.
Especially, it is quite interesting to study the non-Hermitian proximity effects for other systems.
%

We thank Yasuhiro\ Asano, Masatoshi\ Sato, and Kousuke\ Yakubo for helpful discussions.
This work was supported by KAKENHI (Grants
No.\ JP19H00658, No.\ JP20H01828, No.\ JP21H01005, No.\ JP22K03463  and No.\ JP22H01140).

\bibliographystyle{apsrev4-2}
\bibliography{bec_junc}

\end{document}


\title{Supplemental Material for \\ Bulk-Edge Correspondence for Point-Gap Topological Phases in 1D Junction Systems}

\author{Geonhwi Hwang}
\affiliation{Department of Applied Physics, Hokkaido University, Sapporo 060-8628, Japan}
\author{Hideaki Obuse}
\affiliation{Department of Applied Physics, Hokkaido University, Sapporo 060-8628, Japan}
\affiliation{Institute of Industrial Science, The University of Tokyo, 5-1-5 Kashiwanoha, Kashiwa,Chiba 277-8574, Japan}

\renewcommand{\theequation}{S\arabic{equation}}
\setcounter{equation}{0}
\renewcommand{\tablename}{Table S}
\setcounter{table}{0}
\renewcommand{\thefigure}{S\arabic{figure}}
\setcounter{figure}{0}

\begin{abstract}
In this supplemental material, we present the derivation of eigenenergy and eigenfuction in the junction system with periodic and open boundary conditions in a way similar to the non-Bloch band theory.
\end{abstract}

\maketitle
\section{Junction Systems}
\label{S1}
We define the Hamiltonian of the junction system as follows:
%
\begin{equation}
\label{01}
H \coloneqq H_1 (1,N_1) + H_2 (N_1 +1 , N_2-1) +H_\text{BC}.
\end{equation}
Here, $H_{i=1,2}(N,M)$ which corresponds to the subsystem ${\rm I}$ and ${\rm II}$ is given by
%
\begin{align}
\label{02}
H_i (N ,M)\coloneqq \sum_{n=N }^{N + M-1} \left( t_i^{+} c_{n+1}^{\dagger} c_n + t_i^{-} c_n^{\dagger} c_{n+1} + \epsilon_i c_n^{\dagger} c_n  \right) ,
\end{align}
in real space, where $t_i^{\pm} \coloneqq t_i e^{\pm \gamma_i}$, $t_i , \gamma_i , \epsilon_i \in \mathbb{R} $, and $N_i \in \mathbb{N}$ is the size of each subsystem.
For the sake of simplicity, hereafter, we assume $t_i > 0$. 
The third term in the right-hand side (RHS) in Eq.\,(\ref{01}), $H_\text{BC}$, determines the boundary conditions, namely, the periodic boundary conditions(PBC) or open boundary conditions(OBC) as
\begin{align}
 H_\text{BC} &= t_2^+ c_1^\dagger c_{N_1+N_2} + t_2^- c_{N_1+N_2}^\dagger c_1 + \epsilon_2 c_{N_1+N_2}^\dagger c_{N_1+N_2}\quad\quad  (\text{PBC}) \label{eq:PBC},\\
 H_\text{BC} &= \epsilon_2 c_{N_1+N_2}^\dagger c_{N_1+N_2} \quad\quad (\text{OBC})\label{eq:OBC}.
\end{align}

The eigen equation of $H$ is obtained as
%
\begin{equation}
\label{03}
H \ket{\psi} = E \ket{\psi} ,
\end{equation}
where $\ket{\psi} \coloneqq \left(\psi_1 , \cdots , \psi_{N_1+N_2}\right)^T$ is the eigenfunction and $E (\in \mathbb{C})$ is the eigenenergy of the junction system.
Hereafter, we solve the eigen equation in a way similar to the non-Bloch band theory\cite{Yokomizo2019}.

From Eq.\,(\ref{03}), we obtain two recurrence relations for the bulk region as
%
\begin{align}
\label{04}
E_1\psi_n &= t_1^{+} \psi_{n-1} +  t_1^{-} \psi_{n+1} \quad\quad \left(n=2, \cdots , N_1 -1 \right) ,\\
\label{05}
E_2\psi_n &= t_2^{+} \psi_{n-1} +  t_2^{-} \psi_{n+1} \quad\quad \left(n=N_1 +2 , \cdots , N_1+N_2 -1 \right) ,
\end{align}
where $E_1 \coloneqq E- \epsilon_1 $ and $E_2 \coloneqq E- \epsilon_2 $.
Considering the relation at $n=1, N_1, N_1+1,$ and $N_1+N_2$, we obtain four boundary conditions as discussed later.

Without loss of generality, we represent
%
\begin{equation}
\label{06}
E_1 = t_1 \left( x_1 + x_1^{-1} \right), \quad E_2 = t_2 \left( x_2 + x_2^{-1} \right), 
\end{equation}
with $x_1 , x_2 \in \mathbb{C}$ whose absolute values belong to $(0,1]$.
Assuming $\psi_n \propto \lambda^n$, we have the characteristic equations for Eqs.\,(\ref{04}) and (\ref{05}) as
%
\begin{align}
\label{07}
e^{- \gamma_1} {\lambda}^2  - \left( x_1 + x_1^{-1} \right) \lambda + e^{\gamma_1}=0 & \quad\quad \left(n=2, \cdots , N_1 -1 \right) ,\\
\label{08}
e^{- \gamma_2} {\lambda}^2  - \left( x_2 + x_2^{-1} \right) \lambda + e^{\gamma_2}=0 & \quad\quad \left(n=N_1 +2 , \cdots , N_1+N_2 -1 \right).
\end{align}
Consequently, we have the general solutions for the bulk region of each subsystem as a linear combination of $\lambda^n$ for $\lambda$ which satisfies Eqs.\ \ref{07} and \ref{08}, respectively, as
%
\begin{equation}
\label{09}
\psi_n = 
\left\{ \begin{array}{ll}
e^{\gamma_1 n } \left( x_1^n \phi_1 + x_1^{-n} \phi_2 \right) &\quad \left( n=1, \cdots , N_1 \right) ,
\\
e^{\gamma_2 \left( n-N_1\right) } \left( x_2^{n-N_1} \phi_3 + x_2^{-\left(n-N_1 \right)} \phi_4 \right) &\quad \left( n=N_1 +1 , \cdots , N_1 +N_2  \right) ,
\end{array} \right. 
\end{equation}
where $\phi_{i=1,2,3,4}$ are constants.
The equation for the eigenfunction presented above assumes that the values of $x_1$ and $x_2$ are neither $1$ nor $-1$\cite{remark1}.
Note that $x_1$ and $x_2$ are mutually related via
\begin{equation}
\label{eq:spectra}
E= t_1 \left( x_1 + x_1^{-1} \right) +\epsilon_1 = t_2 \left( x_2 + x_2^{-1} \right)+\epsilon_2 . 
\end{equation}

Since we are interested in a continuous spectrum $E$ in the thermodynamic limit, there should be $N_1+N_2$ solutions for $x_1$ and $x_2$.
We denote $m$th solution of a pair of $x_j$ $(j=1,2)$ as $x_j^{(m)}$. In the following two sections, we explain how to obtain such solutions of the junction system with PBC and OBC in Sec.\,\ref{S2} and \ref{S3}, respectively.
%
\section{Junction Systems with PBC}
\label{S2}
Applying Eq.\ (\ref{eq:PBC}) for the boundary condition of the Hamiltonian Eq.\ (\ref{01}), we obtain the four boundary conditions at $n=1,N_1, N_1+1,$ and $N_1+N_2$ satisfying the eigen equation Eq.\ (\ref{03}),
%
%
\begin{align}
\label{1b1}
\left( x_1 + x_1^{-1} \right) \psi_1 &= t e^{\gamma_1 } \psi_{N_1 + N_2 } +  e^{-\gamma_1} \psi_{2} ,
\\
\label{1b2}
\left( x_1 + x_1^{-1} \right) \psi_{N_1} &= e^{\gamma_1} \psi_{N_1 -1} +  e^{-\gamma_1 } \psi_{N_1 +1} ,
\\
\label{1b3}
\left( x_2 + x_2^{-1} \right) \psi_{N_1 +1} &= t^{-1} e^{\gamma_2} \psi_{N_1 } +  e^{-\gamma_2} \psi_{N_1 +2} ,
\\
\label{1b4}
\left( x_2 + x_2^{-1} \right) \psi_{N_1 + N_2 } &= e^{\gamma_2} \psi_{N_1+ N_2 -1 } +  e^{-\gamma_2} \psi_{1} ,
\end{align}
where $t \coloneqq t_2^{+} / t_1^{+}$.
By substituting Eq.\,(\ref{09}) into Eqs.\,(\ref{1b1})-(\ref{1b4}), we have
%
%
%
\begin{equation}
\label{16}
A \ket{\phi} = 0 ,
\end{equation}
where the vector $\ket{\phi} \coloneqq \left(\phi_1 , \phi_2 ,\phi_3 , \phi_4 \right)^T$ and the matrix $A$ is given by 
%
%
\begin{equation}
\label{17}
A = 
\begin{pmatrix}
-1                                                               &-1&
t e^{N_2 \gamma_2 } x_2^{N_2}                            &t e^{N_2 \gamma_2 } x_2^{-N_2}\\
-x_1^{N_1 +1}                                                 &-x_1^{-\left(N_1 +1\right)}&
e^{-\left(N_1 +1\right)\gamma_1} e^{\gamma_2} x_2 &e^{-\left(N_1 +1\right)\gamma_1} e^{\gamma_2} x_2^{-1}\\
t^{-1} e^{N_1 \gamma_1 } x_1^{N_1}                      &t^{-1} e^{N_1 \gamma_1 } x_1^{-N_1}&
-1                                                               &-1\\
e^{-\left(N_2 +1\right)\gamma_2} e^{\gamma_1} x_1 &e^{-\left(N_2 +1\right)\gamma_2} e^{\gamma_1} x_1^{-1}&
-x_2^{N_2 +1}                                                 &-x_2^{-\left(N_2 +1\right)} 
\end{pmatrix} .
\end{equation}
Since the determinant of $A$ must be zero for a nontrivial $\ket{\phi}$, we obtain the following relation
%
%
%
%
%
%
\begin{equation}
\label{111}
\alpha \left[
\cosh\left(N_1 \gamma_1 + N_2 \gamma_2 \right) -\cosh\left(N_1 \log x_1 + N_2 \log x_2 \right) 
\right]
=
\beta \left[
\cosh\left(N_1 \gamma_1 + N_2 \gamma_2 \right) -\cosh\left(N_1 \log x_1 - N_2 \log x_2 \right) 
\right],
\end{equation}
where
\begin{align}
\label{19}
&\alpha\coloneqq \cosh\left[\log\left(t_2 / t_1\right)\right] -\cosh\left[\log\left(x_1 x_2\right)\right] , \\
\label{110}
&\beta\coloneqq \cosh\left[\log\left(t_2 / t_1\right)\right] -\cosh\left[\log\left( x_1 /x_2\right)\right].
\end{align}
Here we consider the case where $N_1=N_2= L $ for simplicity and redefine $x_j \coloneqq e^{-\left(l_j +i \theta_j \right)}$ for $j=1,2$, 
where $l_j  \ge 0$ and $\theta_j \in [0,2\pi)$ by the definition of $x_j$.
Then, for large enough $L$, Eq.\,(\ref{111}) can be approximately expressed as 
%
%
\begin{equation}
\label{112}
\left(\alpha-\beta\right) e^{\left|\gamma_1+\gamma_2\right| L} \approx 
\alpha e^{\left(l_1+l_2\right)L +i\left(\theta_1+\theta_2\right)L}-
\beta e^{\left|l_1-l_2\right|L +i {\rm sgn}(l_1-l_2) \left(\theta_1-\theta_2\right)L}.
\end{equation}

For subsequent calculations, it should be noted that we can only consider the case where $\gamma_1 + \gamma_2 > 0$ without lost of generality.
\subsection{Case I: $l_1, l_2 > 0$}
\label{S2A}
First, we consider the case where $l_1, l_2 >0 $. In this case, the first term in the RHS in Eq.\ (\ref{112}) dominates the second term when $L$ is large enough.
Then, we obtain 
%
%
\begin{equation}
\label{113}
 \frac{\alpha -\beta}{\alpha} \approx \exp\left[\left\{\left(l_1 +l_2 \right)-\left(\gamma_1+\gamma_2\right)\right\}L+i\left(\theta_1 +\theta_2 \right)L\right]. 
\end{equation}
When $l_1+l_2 \ne \gamma_1+\gamma_2$, the RHS diverges or becomes zero for $L \rightarrow \infty$. In this case, the number of the solutions for $x_1$ and $x_2$ is independent on $L$ and then there are no solutions making $E$ continuous.
On the other hand, the case of $l_1+l_2 = \gamma_1+\gamma_2$ gives the condition
\begin{equation}
\label{115}
\frac{\alpha -\beta}{\alpha}  = e^{-i\left(\theta_1 + \theta_2 \right)L},
\end{equation}
which makes the number of the solutions dependent on $L$.
Taking the condition $\l_1 + l_2 = \gamma_1+\gamma_2$ into account, we rewrite the $m$th solution of a pair of $x_j$ ($j=1,2$) as
%
%
\begin{equation}
\label{114}
x_j^{(m)} = e^{-(l_j^{(m)}+i\theta_j^{(m)})},\quad l_j^{(m)} = \gamma_j + (-1)^j \kappa^{(m)},
\end{equation}
where $\kappa^{(m)} (\in \mathbb{R})$ which satisfy
\begin{equation}
\label{eq:caseI_kappa}
-\gamma_2 <\kappa^{(m)}<\gamma_1.
 \end{equation}
While $\theta_1^{(m)}$ and $\theta_2^{(m)}$ have to satisfy Eq.\,(\ref{115}), the condition is quite complicated.
However, we can also obtain the conditions of $\theta_1^{(m)}$ and $\theta_2^{(m)}$ (and $\kappa^{(m)}$) from Eq.\,(\ref{eq:spectra}). Substituting Eq.\,(\ref{114}) into Eq.\,(\ref{eq:spectra}),
we can write down the $m$th eigenenergy $E^{(m)}$ as the function of $\kappa^{(m)}$, $\theta_1^{(m)}$, and $\theta_2^{(m)}$:
%
%
\begin{align}
\label{117}
\begin{aligned}
E^{(m)}&= 2t_1 \cos\left[\theta_1^{(m)} -i\left(\gamma_1 -\kappa^{(m)}\right)\right]+\epsilon_1=2t_1 \left[\cosh\left(\gamma_1 -\kappa^{(m)}\right)\cos\theta_1^{(m)}+i\sinh\left(\gamma_1 -\kappa^{(m)}\right)\sin\theta_1^{(m)}\right] +\epsilon_1
\\
&= 2t_2 \cos\left[\theta_2^{(m)} -i\left(\gamma_2 +\kappa^{(m)}\right)\right]+\epsilon_2=2t_2 \left[\cosh\left(\gamma_2 +\kappa^{(m)}\right)\cos\theta_2^{(m)}+i\sinh\left(\gamma_2 +\kappa^{(m)}\right)\sin\theta_2^{(m)}\right] +\epsilon_2 .
\end{aligned}
\end{align}
As can be seen from Eq.\,(\ref{117}), this case corresponds to the complex spectrum, except for $\theta_j^{(m)}=0,\pi$. Each parameter is mutually connected by Eqs.\,(\ref{115}) and  Eq.\,(\ref{117}) and when a value of one parameter is fixed, the values of the others are simultaneously determined. We will explain the details in Sec.\,\ref{S2D}. By substituting Eq.\,(\ref{114}) into Eq.\,(\ref{09}), the $m$th eigen function corresponding to $E^{(m)}$ is written as
%
%
\begin{equation}
\label{118}
\psi_n^{(m)} \propto 
\left\{ \begin{array}{ll}
e^{\gamma_1 n}  \sinh\left[\left(\gamma_1 -\kappa^{(m)} +i\theta_1^{(m)} \right) \left(n-L\right)\right] &\quad \left( n=1, \cdots , L \right) ,
\\
\\
e^{\gamma_2 n}  \sinh\left[\left(\gamma_2 +\kappa^{(m)} +i\theta_2^{(m)} \right) \left(n-2L\right)\right] &\quad \left( n=L +1 , \cdots , 2L  \right) .
\end{array} \right. 
\end{equation}
Apart from the interfaces at $n=1$ and $L$, the following approximation is nearly satisfied:
%
%
\begin{equation}
\label{119}
\left| \psi_n^{(m)} \right| \propto 
\left\{ \begin{array}{ll}
\exp(\kappa^{(m)} n)  &\quad \left( n=1, \cdots , L \right) ,
\\
\exp(-\kappa^{(m)} n) &\quad \left( n=L +1 , \cdots , 2L  \right) .
\end{array} \right. 
\end{equation}
Taking the above result into account, it is reasonable to call $\kappa^{(m)}$ the inverse of the localization length.
\subsection{Case II: $l_1 =0$, $l_2 > 0$.}
\label{S2B}
Here, we consider the case where $l_1=0$ and $l_2 >0 $. 
First, we remark that the eigenenergy $E$ in Case II should be real as $E_1=2t_1 \cos \theta_1 \in \mathbb{R}$ due to $l_1=0$. 

Considering $l_1=0$ and $l_2 >0 $, Eq.\,(\ref{112}) is written down as
%
%
\begin{equation}
\label{120}
\frac{\alpha -\beta}{\alpha e^{i\left(\theta_1 +\theta_2 \right)L} -\beta e^{i\left(- \theta_1 +\theta_2\right)L}} \approx e^{\left( l_2 - \left(\gamma_1 +\gamma_2 \right)\right)L} . 
\end{equation}

First, we consider the case where $l_2 < \gamma_1 + \gamma_2$. In this case, the RHS in Eq.\,(\ref{120}) becomes zero in the limit of $L\rightarrow \infty$, and then
the numerator in the left-hand side (LHS) must be zero, namely, $\alpha=\beta$. According to the definition of $\alpha$ and $\beta$ in Eqs.\,(\ref{19}) and (\ref{110}) and the conditions $l_1=0$ and $l_2>0$, 
$\alpha=\beta$ is satisfied by
%
%
\begin{equation}
\label{121}
l_1 = 0 ,\quad \theta_1= 0 , \pi.
\end{equation}
However, this condition makes the LHS finite at the same time since the denominator also becomes zero. Consequently, no solution exists where $l_2 <\gamma_1 +\gamma_2$.

Next, we consider the opposite condition, $l_2 > \gamma_1 +\gamma_2$, which makes the RHS in Eq.\,(\ref{120}) diverge when $L \rightarrow \infty$.
This requires that the denominator in the LHS becomes zero.
Then, from Eq.\,(\ref{120}), the condition for solution to exist is given by
%
%
\begin{equation}
\label{122}
\frac \beta \alpha = e^{i2\theta_1 L}\quad\quad (l_2 > \gamma_1 +\gamma_2) ,
\end{equation}
which makes the number of the solutions dependent on $L$.

Finally, for $l_2=\gamma_1+\gamma_2$,
\begin{equation}
\label{122b}
\frac \beta \alpha =\frac{\sin[(\theta_1+\theta_2)L/2]}{\sin[(-\theta_1+\theta_2)L/2]} e^{i\theta_1 L}\quad\quad (l_2= \gamma_1 +\gamma_2) .
\end{equation}
Therefore, the number of the solutions depends on $L$ for $l_2 \ge \gamma_1+\gamma_2$.

In the same way as in Sec.\,\ref{S2A}, from Eq.\,(\ref{eq:spectra}), we can describe the $m$th eigenenergy $E^{(m)}$ as the function of $l_2^{(m)}$, $\theta_1^{(m)}$, and $\theta_2^{(m)}$:
%
\begin{equation}
\label{123}
\begin{aligned}
E^{(m)} &= 2t_1 \cos\theta_1^{(m)} +\epsilon_1\\
&= 2t_2 \cos(\theta_2^{(m)}-il_2^{(m)}) +\epsilon_2.
\end{aligned}
\end{equation}
This again gives solutions as
\begin{equation}
l_1^{(m)}=0,\quad l_2^{(m)}\geq \gamma_1+\gamma_2,\quad \theta_1^{(m)} \in [0,2\pi) , \quad \theta_2^{(m)}=0,\pi ,
\label{eq:sol-case2}
\end{equation}
where the specific values of $l_2^{(m)}$, $\theta_1^{(m)}$, and $\theta_2^{(m)}$ are determined so that they satisfy Eqs.\,(\ref{122}) and (\ref{123}) for $l_2>\gamma_1+\gamma_2$ [or Eqs.\,(\ref{122b}) and (\ref{123}) for $l_2=\gamma_1+\gamma_2$].
We will explain the details in Sec.\,\ref{S2D}.

Regarding the eigenfunctions, the following is approximately satisfied:
%
%
\begin{equation}
\label{124}
\left| \psi_n^{(m)} \right| \propto 
\left\{ \begin{array}{ll}
\exp(\gamma_1 n) &\quad \left( n=1, \cdots , L \right) ,
\\
\exp\left[ \left(\gamma_2 - {\rm sgn}(\gamma_1)l_2^{(m)}  \right) n\right] &\quad \left( n=L +1 , \cdots , 2L  \right) .
\end{array} \right. 
\end{equation}
We see that the localization length is different in both subsystems for $l_2>\gamma_1 + \gamma_2$.

We can also consider the case where $l_1>0$ and $l_2=0$ in the same way.
\subsection{Case III: $l_1= l_2 = 0$.}
\label{S2C}
In this case, Eq.\,(\ref{111}) is written down as
\begin{equation}
\label{eq:l0}
\frac{\alpha -\beta}{\alpha \cos [\left(\theta_1 +\theta_2 \right) L] -\beta \cos [\left(- \theta_1 +\theta_2\right) L]} \approx 2e^{-\left( \gamma_1 +\gamma_2 \right)L} . 
\end{equation}
As in Sec.\,\ref{S2B}, the numerator in the LHS must be zero in the limit of $L\rightarrow \infty$, which makes the LHS finite at the same time.
Thus, no solution exists where $l_1=l_2=0$.
\subsection{Bulk-Edge Correspondence for Junction Systems with PBC}
\label{S2D}
In the main text, we claim that the spectra for junction systems with PBC, $\sigma_{\rm PBC}^{\rm junc}$, do not appear where $\Delta w= 0$, namely, where the winding number is equal for both subsystems, which is also confirmed numerically as shown in Figs.\,1 and 3(a) in the main text.
Here, we explain how to reach at the bulk-edge correspondence (BEC) for the point-gap topological phases in junction systems with PBC from the results presented above.

First of all, we emphasize that we use the two conditions to determine the eigenenergies and eigenstates of the junction system with PBC, namely, Eq.\,(\ref{eq:spectra}) and the condition derived from the nontrivial solution of $\ket{\psi}$ in Eq.\,(\ref{16}).
We will explain the details hereafter.

{\it In Case I in Sec.\,\ref{S2A}}:
\begin{figure}
  \centering
  \includegraphics[width=\linewidth]{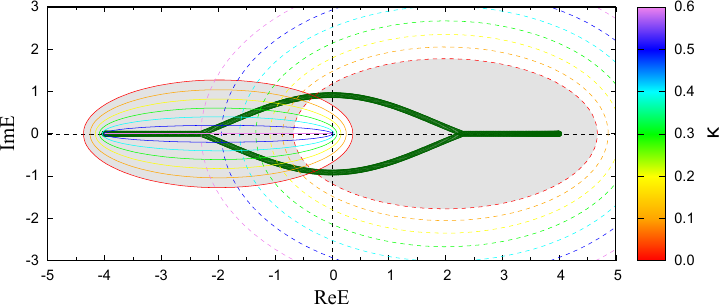}
\caption{Spectrum of a junction system with PBC (green dots) with parameters $t_1=t_2=1$, $\gamma_1=0.6$, $\gamma_2=0.8$, $\epsilon_1=-\epsilon_2=-2.0$, and $N_1=N_2=500$.
Two ellipses $E_1^{(m)}+\epsilon_1$(solid lines) and $E_2^{(m)}+\epsilon_2$(dashed lines) are shown by the color (red to violet) corresponding to the value of $\kappa$ ($0$ to $0.6$ at $0.1$ intervals).
The gray region means $\Delta w \neq 0$.}
\label{f1}
\end{figure}
We use the relations in Eqs.\,(\ref{113}) and (\ref{117}) [The relation Eq.\,(\ref{eq:caseI_kappa}) is derived from Eq.\,(\ref{113})]. 
Figure\,\ref{f1} demonstrates how to decide the parameters $\kappa^{(m)}$, $\theta_1^{(m)}$, and $\theta_2^{(m)}$ by considering the junction system in Fig.\,$1$(c) in the main text, for example.
The solid (dashed) curve in Fig.\,\ref{f1} represents an ellipse drawn from $E_1^{(m)}+\epsilon_1$ ($E_2^{(m)}+\epsilon_2$) with a fixed value of $\kappa^{(m)}$ as a function of $\theta_1^{(m)}$ ($\theta_2^{(m)}$) $\in [0,2\pi)$. At a certain range of $\kappa^{(m)}$, the two ellipses may overlap each other and there appear the two crossing points. In that case, we can determine $\theta_1^{(m)}$ and $\theta_2^{(m)}$ which indicate the crossing points. As varying the value of $\kappa^{(m)}$ in the range, we can obtain the solutions $\kappa^{(m)}$ and $\theta_j^{(m)}$ ($j=1,2$) and then eigenenergies $E^{(m)}$ which becomes continuous spectrum in the thermodynamic limit.

If the winding number is the same in both subsystems, {\it e.g.}  $\gamma_1\gamma_2 >0$, the crossing points could appear at $\kappa^{(m)}=0$ since the range in Eq.\,(\ref{eq:caseI_kappa}) includes $\kappa^{(m)}=0$. However, this also depends on the parameters of the model. The crossing points appear at $\kappa^{(m)}=0$ for the system shown in Fig.\,\ref{f1}, while the crossing points do not appear at $\kappa^{(m)}=0$ but appear at a ceretain value $\kappa^{(m)}$ in the range of Eq.\,(\ref{eq:caseI_kappa}) in the case of Fig.\,$1$(b) in the main text. Further, In the case of Fig.\,$1$(a) in the main text, the crossing points never occurs in the range in Eq.\,(\ref{eq:caseI_kappa}).
We remark that $\kappa^{(m)}=0$ is not allowed where $\gamma_1 \gamma_2 < 0$ from Eq.\,(\ref{eq:caseI_kappa}).

When $\kappa^{(m)}=0$, the ellipse $E_j^{(m)}+\epsilon_j$ is identical with the spectrum for the corresponding subsystem with PBC, $\sigma_{\rm PBC}^{(j)}$, for $j=1,2$.
Thereby, appearing two crossing points at $\kappa^{(m)}=0$ ($l_j^{(m)}=\gamma_j$ for $j=1,2$) means that the spectra for both subsystems with PBC intersects and there exists the region with $\Delta w=0$.
If $\sigma_{\rm PBC}^{(1)}$ and $\sigma_{\rm PBC}^{(2)}$ overlap, as in Fig.\,1(c) in the main text, there always be two eigenenergies at the crossing points of $\sigma_{\rm PBC}^{(1)}$ and $\sigma_{\rm PBC}^{(2)}$ which can be derived by substituting $\kappa^{(m)}=0$ into Eq.\,(\ref{117}).
Varying the value of $\kappa^{(m)}$ from zero to $\gamma_1$ ($-\gamma_2$), the ellipse $E_1^{(m)}+\epsilon_1$ gets smaller (larger) and the other ellipse $E_2^{(m)}+\epsilon_2$ gets larger (smaller) from the spectrum for each subsystem with PBC. As indicated in Fig.\,\ref{f1}, the crossing points of two ellipses do not appear in the region where $\Delta w=0$.
When the value of $\kappa^{(m)}$ approaches the maximum or minimum value, one ellipse almost collapses on the real axis and becomes almost the same with $\sigma_{\rm OBC}^{(j=1,2)}$, the spectrum for the corresponding subsystem with OBC.
Even if there are no crossing point at $\kappa^{(m)}=0$, we can consider where the complex eigenenergies of junction systems with PBC appears in the same way.
Given that the two ellipses never get smaller or larger simultaneously with continuously varying $\kappa^{(m)}$, we know that the complex eigenenergies $E^{(m)}$ must appear on the region where $\Delta w \neq 0$, together with $E^{(m)} \in(\sigma_{\rm PBC}^{(1)} \cap \sigma_{\rm PBC}^{(2)})$ for junction systems with $\gamma_1 \gamma_2 >0$ where $\sigma_{\rm PBC}^{(1)}$ and $\sigma_{\rm PBC}^{(2)}$ overlap.
This proves that all the complex eigenenergies of junction systems satisfy Eq.\,($19$) in the main text.

{\it In Case II in Sec.\,\ref{S2B}}:
\begin{figure}
  \centering
  \includegraphics[width=\linewidth]{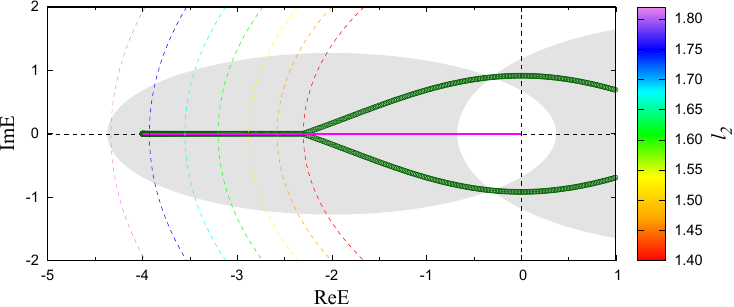}
\caption{Spectrum of a junction system with PBC (green dots) with parameters $t_1=t_2=1$, $\gamma_1=0.6$, $\gamma_2=0.8$, $\epsilon_1=-\epsilon_2=-2.0$, and $N_1=N_2=500$.
Line segment $E_1^{(m)}+\epsilon_1$ is shown in magenta line and ellipse $E_2^{(m)}+\epsilon_2$(dashed lines) is shown by the color (red to violet) corresponding to the value of $l_2$ ($1.4$ to $1.82$ at $0.07$ intervals).
The gray region means $\Delta w \neq 0$.}
\label{f2}
\end{figure}
We use the relations in Eqs.\,(\ref{120}) and (\ref{123}) [The relation Eqs.\,(\ref{122}) and (\ref{122b}) are derived from Eq.\,(\ref{120})]. 
Figure\,\ref{f2} demonstrates how to decide the parameters $l_2^{(m)}$, $\theta_1^{(m)}$, and $\theta_2^{(m)}$ by considering the junction system in Fig.\,$1$(c) in the main text, for example.
The magenta solid line in Fig.\,\ref{f2} represents $E_1^{(m)}+\epsilon_1$, which is identical with $\sigma_{\rm OBC}^{(1)}$.
The dashed curve represents an ellipse drawn from $E_2^{(m)}+\epsilon_2$ with a fixed value of $l_2^{(m)}$ as a function of $\theta_2^{(m)} \in [0,2\pi)$.
At a certain range of $l_2^{(m)}$, $E_1^{(m)}+\epsilon_1 \in \sigma_{\rm OBC}^{(1)}$ and $E_2^{(m)}+\epsilon_2$ may overlap each other and there appears the one crossing point.
In that case, we can determine the $\theta_1^{(m)}$ and $\theta_2^{(m)}$ which indicate the crossing point.
As varying the value of $l_2^{(m)}$ in the range, we can obtain the solutions $l_2^{(m)}$ and $\theta_j^{(m)}$ ($j=1,2$) and then eigenenergy $E^{(m)}$ which becomes continuous spectrum in the thermodynamic limit.

As mentioned above, the condition $l_1=0$ forces $E_1+\epsilon_1=2\cos \theta_1+\epsilon_1$ which is identical with $\sigma_{\rm OBC}^{(1)}$.
We can also consider the case where $l_1>0$ and $l_2=0$.
In this case, the condition $l_2=0$ forces $E_2+\epsilon_2=2\cos \theta_2+\epsilon_2$ which is identical with $\sigma_{\rm OBC}^{(2)}$.
This means that all the eigenenergies of junction systems with PBC corresponding to Case II should be real, and even should appear on $\sigma_{\rm OBC}^{(1)}$ or $\sigma_{\rm OBC}^{(2)}$.
Finally, we know that there is no solution in Case III.
Then, we can conclude that all the real eigenenergies which satisfy $l_1 l_2=0$ should appear on $\sigma_{\rm OBC}^{(1)}$ or $\sigma_{\rm OBC}^{(2)}$.
This means that these real eigenenergies never appear outside $\sigma_{\rm PBC}^{(1)}$ and $\sigma_{\rm PBC}^{(2)}$ by applying the existing BEC for point-gap topological phases in systems with PBC and OBC\cite{Gong2018,Okuma2020,Zhang2020}.

However, the above statement is not enough to reach at the BEC for the point-gap topological phases in junction systems with PBC.
Because, if $\gamma_1 \gamma_2>0$, there may be a region where $\Delta w=0$ even inside $\sigma_{\rm PBC}^{(1)}$ and $\sigma_{\rm PBC}^{(2)}$ (say region $S$) depending on the parameters, as the non-gray region shared by $\sigma_{\rm PBC}^{(1)}$ and $\sigma_{\rm PBC}^{(2)}$ in Fig.\,\ref{f2}.
Then, there remains a possibility that $\sigma_{\rm OBC}^{(j=1,2)}$ and the region $S$ overlap.
To confirm whether the BEC for the point-gap topological phases in junction systems with PBC is valid even when the region $S$ exists, we explain as follows.
$l_1=0$ which makes $E_1+\epsilon_1 \in \sigma_{\rm OBC}^{(1)}$ occurs only in Case II.
If the eigenenergies for the junction system with PBC appear in the region $S$, $l_2$ should be smaller than $\gamma_2$.
However, in Case II, $l_2 \ge \gamma_1+\gamma_2$ is required for the continuous spectrum. Therefore, the eigenenergies never appear in the region $S$.
We can also explain the case $l_1>0$ and $l_2=0$ in the same way.
Therefore, we prove that all the real eigenenergies of junction systems satisfy Eq.\,($19$) in the main text.

Here is the summary for the above.
First, every eigenenergy of junction systems with PBC is given by the crossing points of $E_1+\epsilon_1$ and $E_2+\epsilon_2$.
Second, for all the cases, we obtain a parameter which determines the length of both $x_1$ and $x_2$ simultaneously and then we can consider $E_1+\epsilon_1$ and $E_2+\epsilon_2$ as ellipses.
Third, with the continuously varying parameter, one ellipse gets smaller and the other gets larger and finally the former one collapses on $\sigma_{\rm OBC}^{(1)}$ or $\sigma_{\rm OBC}^{(2)}$, with the other keeps getting larger.

Therefore, we understand that $\sigma_{\rm PBC}^{\rm junc}$ appears on and inside $\sigma_{\rm PBC}^{(1)}$ or $\sigma_{\rm PBC}^{(2)}$ where $\Delta w \ne 0$ as confirming analytically.
This establishes the bulk-edge correspondence for junction systems with PBC.

{\it Non-Hermitian proximity effect}:
There are three remarks on the eigenfunctions.
First, according to Eqs.\,(\ref{119}) and (\ref{124}), the peaks of the wavefunction $|\psi_n|^2$ appears near the interface at $n=1$ (or $L$). Note that the peak appears to satisfy ``continuity'' of wavefunction. If the peak appears at the left(right) edge of the subsystem I, the peak appears at the right(left) edge of the subsystem II and vice versa. We call this phenomenon non-Hermitian proximity effects as we explained in the main text. 
Second, the localization length is the same in both subsystem I and II in Case I where the corresponding eigenenergy is complex, while $\gamma_1 \ne \gamma_2$ in general. This is the one of the interesting properties of non-Hermitian proximity effects.
Third, the localization length diverges if $\kappa^{(m)}=0$. This occurs only when the corresponding eigenenergy $E^{(m)}$ of the junction system with PBC appears at the crossing points of the spectra of both subsystems with PBC, $\sigma_{\rm PBC}^{(1)}$ and $\sigma_{\rm PBC}^{(2)}$.

%
%
%
%
%
\section{Junction Systems with OBC}
\label{S3}
While there are $N_1+N_2$ cases to implement OBC by cutting the hopping between two neighboring sites in the junction system with PBC, we focus on a junction system with OBC applying Eq.\,(\ref{eq:OBC}) to Eq.\,(\ref{01}) for simplicity.
%
%
%
In the same way as in Sec.\,\ref{S2}, we derive the four boundary conditions:
%
\begin{align}
\label{2b1}
\left( x_1 + x_1^{-1} \right) \psi_1 &=  e^{-\gamma_1} \psi_{2},
\\
\label{2b2}
\left( x_1 + x_1^{-1} \right) \psi_{N_1} &= e^{\gamma_1} \psi_{N_1 -1} +  e^{-\gamma_1 } \psi_{N_1 +1},
\\
\label{2b3}
\left( x_2 + x_2^{-1} \right) \psi_{N_1 +1} &= t^{-1} e^{\gamma_2} \psi_{N_1 } +  e^{-\gamma_2} \psi_{N_1 +2},
\\
\label{2b4}
\left( x_2 + x_2^{-1} \right) \psi_{N_1 + N_2 } &= e^{\gamma_2} \psi_{N_1 +N_2 -1 } .
\end{align}
By substituting Eq.\,(\ref{09}) into Eqs.\,(\ref{2b1})-(\ref{2b4}), we have
%
%
%
\begin{equation}
\label{26}
A \ket{\phi} = 0 ,
\end{equation}
where the vector $\ket{\phi} \coloneqq \left(\phi_1 , \phi_2 ,\phi_3 , \phi_4 \right)^T$ and the matrix $A$ is given by 
%
%
%
%
\begin{equation}
\label{27}
A = 
\begin{pmatrix}
-1                                                               &-1&
0                                                                 &0 \\
-x_1^{N_1 +1}                                                 &-x_1^{-\left(N_1 +1\right)}&
e^{-\left(N_1 +1\right)\gamma_1} e^{\gamma_2} x_2 &e^{-\left(N_1 +1\right)\gamma_1} e^{\gamma_2} x_2^{-1}\\
t^{-1} e^{N_1 \gamma_1 } x_1^{N_1}                      &t^{-1} e^{N_1 \gamma_1 } x_1^{-N_1}&
-1                                                               &-1\\
0                                                                 &0 &
-x_2^{N_2 +1}                                                 &-x_2^{-\left(N_2 +1\right)} 
\end{pmatrix}.
\end{equation}
Since the determinant of $A$ must be zero for a nontrivial $\ket{\phi}$, we obtain the following relation:
%
%
\begin{equation}
\label{28}
\left(x_1^{N_1+1} -x_1^{-\left(N_1+1\right)} \right) \left(x_2^{N_2+1} -x_2^{-\left(N_2+1\right)} \right)
= t^{-1} e^{\gamma_2 -\gamma_1} \left(x_1^{N_1} -x_1^{-N_1} \right) \left(x_2^{N_2} -x_2^{-N_2} \right).
\end{equation}
Here we consider the case where $N_1=N_2= L $ for simplicity and redefine $x_j \coloneqq e^{-\left(l_j +i \theta_j \right)}$ for $j=1,2$, where $l_j  \ge 0$ and $\theta_j \in [0,2\pi)$ by the definition of $x_j$.
Then, Eq.\,(\ref{28}) is written down as
\begin{equation}
\label{29}
t_1 \sinh\left[\left(l_1 +i\theta_1\right)L\right] \sinh\left[\left(l_2 +i\theta_2\right)L\right] =
t_2 \sinh\left[\left(l_1 +i\theta_1\right)\left(L+1\right)\right] \sinh\left[\left(l_2 +i\theta_2\right)\left(L+1\right)\right] .
\end{equation}
In a similar way as in Sec.\,\ref{S2A} and Sec.\,\ref{S2B}, the eigenfunction can be approximated as
%
%
\begin{equation}
\label{20}
\left| \psi_n \right| \propto 
\left\{ \begin{array}{ll}
\exp[(\gamma_1 + l_1 ) n] &\quad \left( n=1, \cdots , L \right) ,
\\
\exp[(\gamma_2 - l_2 ) n] &\quad \left( n=L +1 , \cdots , 2L  \right) .
\end{array} \right. 
\end{equation}
\subsection{Case I: $l_1, l_2 > 0$}
\label{S3A}

First, we consider the case where $l_1, l_2 >0 $.
For large enough $L$, Eq.\,(\ref{29}) can be approximately expressed as 
\begin{equation}
\label{210}
\frac{t_1}{t_2} \approx \exp \left[(l_1+l_2) +i(\theta_1+\theta_2)\right].
\end{equation}
From Eq.\,(\ref{210}), we see that the number of the solutions for $x_1$ and $x_2$ is independent on $L$.
Therefore, in Case I, there are no solutions making $E$ continuous.
\subsection{Case II: $l_1 =0$, $l_2 \geq 0$.}
\label{S3B}
Here, we consider the case where $l_1=0$ and $l_2 \geq 0 $. First, for $l_2>0$, Eq.\,(\ref{29}) can be approximately expressed as 
\begin{equation}
\label{212}
\frac{t_1 \sin [\theta_1 L]}{t_2 \sin [\theta_1 (L+1)]} \approx \exp \left(l_2 +i\theta_2\right) \quad\quad (l_2>0).
\end{equation}
Second, for $l_2=0$, Eq.\,(\ref{29}) is written down as
\begin{equation}
\label{214}
\frac{t_1 \sin [\theta_1 L]\sin [\theta_2 L]}{t_2 \sin [\theta_1 (L+1)]\sin [\theta_2 (L+1)]}=1\quad\quad (l_2=0).
\end{equation}
Therefore, the number of the solutions depends on $L$ for $l_2 \geq 0$.

In the same way as in Sec.\,\ref{S2B}, from Eq.\,(\ref{eq:spectra}), we can describe the $m$th eigenenergy $E^{(m)}$ as the function of $l_2^{(m)}$, $\theta_1^{(m)}$, and $\theta_2^{(m)}$:
%
\begin{equation}
\label{215}
\begin{aligned}
E^{(m)} &= 2t_1 \cos\theta_1^{(m)} +\epsilon_1\\
&= 2t_2 \cos(\theta_2^{(m)}-il_2^{(m)}) +\epsilon_2.
\end{aligned}
\end{equation}
This again gives solutions as
\begin{equation}
l_1^{(m)}=0,\quad l_2^{(m)}\geq 0,\quad \theta_1^{(m)}\in[0,2\pi) , \quad \theta_2^{(m)}=0,\pi , 
\label{eq:sol-case2}
\end{equation}
where the specific values of $l_2^{(m)}$, $\theta_1^{(m)}$, and $\theta_2^{(m)}$ are determined so that they satisfy Eqs.\,(\ref{212}) and (\ref{215}) for $l_2>0$ [or Eqs.\,(\ref{214}) and (\ref{215}) for $l_2=0$].
We will explain the details in Sec.\,\ref{S3D}. 
Substituting $l_1^{(m)}=0$ and $l_2^{(m)}>0$ into Eq.\,(\ref{20}), we can derive the corresponding eigenfunction.

We can also consider the case where $l_1>0$ and $l_2=0$ in the same way.
%
%
%
%
%
%
\subsection{Bulk-Edge Correspondence for Junction Systems with OBC}
\label{S3D}
In the main text, we claim that the spectra for the junction systems with OBC, $\sigma_{\rm OBC}^{\rm junc}$, do not appear where $w^{\rm junc}= 0$, namely, where the winding number of the junction system is zero, which is also confirmed numerically as shown in Fig.\,4(b) in the main text.
Here, we explain how to reach at the BEC for the point-gap topological phases in junction systems with OBC from the results presented above.

We use the relations in Eq.\,(\ref{212}) and (\ref{215}) for $l_2>0$ [or Eq.\,(\ref{214}) and (\ref{215}) for $l_2=0$]. 
We know how to decide the parameters $l_2^{(m)}$, $\theta_1^{(m)}$, and $\theta_2^{(m)}$ if $l_1=0$ and $l_2> 0$ as demonstrated in Fig\,\ref{f2}.
Only the difference from Case II in the Sec.\,\ref{S2B} is the lower bound, $l_2 \ge \gamma_1+\gamma_2$ for PBC or $l_2\ge 0$ for OBC.
At a certain range of $l_2^{(m)}$, $E_1^{(m)}+\epsilon_1 \in \sigma_{\rm OBC}^{(1)}$ and $E_2^{(m)}+\epsilon_2$ may overlap each other and there appears the one crossing point.
In that case, we can determine the $\theta_1^{(m)}$ and $\theta_2^{(m)}$ which indicate the crossing point.
As varying the value of $l_2^{(m)}$ in the range, we can obtain the solutions $l_2^{(m)}$ and $\theta_j^{(m)}$ ($j=1,2$) and then eigenenergy $E^{(m)}$ which becomes continuous spectrum in the thermodynamic limit.

We know the condition $l_1=0$ forces $E_1+\epsilon_1=2\cos \theta_1+\epsilon_1$ which is identical with $\sigma_{\rm OBC}^{(1)}$.
We can also consider the case where $l_1 \geq 0$ and $l_2=0$; the condition $l_2=0$ forces $E_2+\epsilon_2=2\cos \theta_2+\epsilon_2$ which is identical with $\sigma_{\rm OBC}^{(2)}$.
Finally, we know that there is no solution in Case I.
Therefore, we can conclude that all the eigenenergies should appear on $\sigma_{\rm OBC}^{(1)}$ or $\sigma_{\rm OBC}^{(2)}$.

Meanwhile, in Sec.\,\ref{S2D}, we can see that $\sigma_{\rm OBC}^{(1)}$ and $\sigma_{\rm OBC}^{(2)}$ are always included within or inside $\sigma_{\rm PBC}^{\rm junc}$.
Consequently, $\sigma_{\rm OBC}^{\rm junc}$ is always included within or inside $\sigma_{\rm PBC}^{\rm junc}$.
This proves that all the spectra of junction systems with OBC satisfy the following equation:
\begin{align}
\label{eq:BEC}
\sigma_{\rm OBC}^{\rm junc} \subset \{ E \vert w^{\rm junc} (E) \neq 0 \lor E \in \sigma_{\rm PBC}^{\rm junc} \} .
\end{align}

Therefore, we understand that $\sigma_{\rm OBC}^{\rm junc}$ appears on and inside $\sigma_{\rm PBC}^{\rm junc}$ where $w^{\rm junc} \ne 0$ as confirming analytically.
This establishes the bulk-edge correspondence for junction systems with OBC.

Regarding the eigenfunction, if $l_1=0$ and $l_2>\gamma_2$, the peak of the wavefunction $|\psi_n|^2$ appears near $n=L$ in both subsystems to satisfy ``continuity'' of the wavefunction. This is the same with the non-Hermitian proximity effects as we mentioned in Sec.\,\ref{S2D}.
However, in the case of OBC, the peak of $|\psi_n|^2$ could appear only at the right edge in the subsystem II as shown in Fig.\,4(b) in the main paper. 
This is because the OBC forbade the proximity effects between $x=1$ and $x=2L$.